\documentclass[aip, apl, reprint]{revtex4-2} % APL Photonics official document class

\usepackage{graphicx}
\usepackage{amsmath}
\usepackage{amssymb}
\usepackage{orcidlink}
\usepackage{xcolor}

\begin{document}
	
	\title{Quantitative Dynamic Phase Mapping via Single-Arm Field-Correlation Ghost Imaging}
	
	\author{Chaoran Wang~\orcidlink{0009-0009-6585-7199}}
	\email{chaoranwang@siom.ac.cn}
	\affiliation{Aerospace Laser Technology and System Department, Shanghai Institute of Optics and Fine Mechanics, Chinese Academy of Sciences, Shanghai 201800, China}
	\affiliation{Center of Materials Science and Optoelectronics Engineering, University of Chinese Academy of Sciences, Beijing 100049, China}
	
	\author{Jinquan Qi~\orcidlink{0009-0004-7183-4210}}
	\affiliation{Aerospace Laser Technology and System Department, Shanghai Institute of Optics and Fine Mechanics, Chinese Academy of Sciences, Shanghai 201800, China}
	\affiliation{Center of Materials Science and Optoelectronics Engineering, University of Chinese Academy of Sciences, Beijing 100049, China}
	
	\author{Shuang Liu~\orcidlink{0000-0003-2504-2349}}
	\affiliation{Aerospace Laser Technology and System Department, Shanghai Institute of Optics and Fine Mechanics, Chinese Academy of Sciences, Shanghai 201800, China}
	
	\author{Xingzhao Jiang~\orcidlink{0009-0007-0040-9358}}
	\affiliation{Key Laboratory of Materials Modification by Laser, Ion, and Electron Beams (Ministry of Education), School of Physics, Dalian University of Technology, Dalian 116024, China}
	
	\author{Shensheng Han~\orcidlink{0000-0003-1689-5876}}
	\email{sshan@mail.shcnc.ac.cn}
	\affiliation{Aerospace Laser Technology and System Department, Shanghai Institute of Optics and Fine Mechanics, Chinese Academy of Sciences, Shanghai 201800, China}
	\affiliation{Center of Materials Science and Optoelectronics Engineering, University of Chinese Academy of Sciences, Beijing 100049, China}
	\affiliation{Hangzhou Institute for Advanced Study, University of Chinese Academy of Sciences, Hangzhou 310024, Zhejiang, China}

	\begin{abstract}
		We demonstrate a single-arm optical platform for phase-retrieval-free, quantitative dynamic phase mapping of continuous transparent media via field-correlation ghost imaging. By modeling the medium as a dynamic pure-phase object, we spatially encode and compress its two-dimensional (2D) complex transmittance into a single bucket detector. Balanced heterodyne detection downconverts the optical frequencies for direct digitization. By mapping spatial information into the temporal domain, this single-pixel architecture utilizes high-speed digitization to continuously resolve 2D phase dynamics, overcoming the frame-rate bottlenecks of traditional array sensors. Coupled with intermediate frequency spectral analysis, this establishes a direct linear mapping from the recorded signal to the physical phase. The complex amplitude is thus deterministically extracted via field-correlation, enabling the spatial reconstruction of 2D acoustic pressure distributions using a pseudo-inverse algorithm. Experimental validations in an acoustic levitator confirm that the optically extracted acoustic wavelengths closely match theoretical dispersion models, exhibiting a strong linear correlation between the retrieved phase shift and local sound pressure levels. This deterministic methodology provides a real-time metrological tool for characterizing rapidly evolving phenomena, including transient aeroacoustic flows, shockwaves, and microfluidic biological dynamics.
	\end{abstract}
	
	\maketitle % In REVTeX, maketitle must be placed AFTER the abstract
	
	\section{Introduction}
	
	In electromagnetic wave measurements, phase is inherently a more sensitive parameter than amplitude across the microwave \cite{nikolovaMicrowaveImagingBreast2011}, terahertz (THz) \cite{tonouchiCuttingedgeTerahertzTechnology2007}, and X-ray \cite{pfeifferPhaseRetrievalDifferential2006, yuFouriertransformGhostImaging2016} regimes. Traditional dual-arm techniques, such as holography \cite{cucheDigitalHolographyQuantitative1999, huangQuantitativePhaseImaging2024} and interferometry \cite{smithTurbulencefreeDoubleslitInterferometer2018}, offer high sensitivity but demand stringent environmental stability, making them highly susceptible to mechanical vibrations and optical path length drifts.
	
	To mitigate these stability constraints, non-interferometric single-arm techniques—such as coherent diffractive imaging (CDI) \cite{fienupPhaseRetrievalAlgorithms1982, parkQuantitativePhaseImaging2018}, the transport of intensity equation (TIE) \cite{huangQuantitativePhaseImaging2024}, diffraction phase microscopy \cite{popescuDiffractionPhaseMicroscopy2006}, and ptychography \cite{zhengWidefieldHighresolutionFourier2013}—are widely used. These methods record two-dimensional (2D) diffraction intensities and employ iterative, constraint-based algorithms to retrieve the phase of the optical field \cite{fienupPhaseRetrievalAlgorithms1982}. However, these computational retrieval processes demand a high signal-to-noise ratio (SNR). Furthermore, the low temporal sampling rates of 2D array detectors—even for modern SPADs and lock-in cameras \cite{bruschiniSinglephotonAvalancheDiode2019,foixLockinTimeofflightToF2011}—restrict these techniques to time-averaged or single-shot transient captures, hindering continuous dynamic phase imaging. Alternative high-speed schemes face similar limitations  \cite{liangSingleshotRealtimeVideo2017}, pump-probe imaging requires strict phase-locking, while compressed ultrafast sampling methods \cite{liangSingleshotStereopolarimetricCompressed2020} predominantly capture intensity rather than phase.
	
	Ghost imaging (GI) \cite{shapiro2008computational, erkmen2010ghost} provides a robust, single-arm alternative that significantly enhances stability against environmental perturbations \cite{meyers2011turbulence,ilinaAberrationinsensitiveMicroscopyUsing2019}. By probing the target with a single arm devoid of spatially resolving detectors \cite{duarteSinglepixelImagingCompressive2008, edgarPrinciplesProspectsSinglepixel2019, clemente2010optical}, GI achieves high resilience, even when operating through scattering layers \cite{katz2014noninvasive}. To enable phase measurement while preserving single-arm robustness, coherent detection computational ghost imaging (CD-GI) incorporates heterodyne or homodyne detection \cite{bacheGhostImagingUsing2004, zhangHomodyneDetectionGhost2009} to recover both the amplitude and phase of the transmitted light \cite{gongPhaseretrievalGhostImaging2010,wang2012coherent}. Crucially, its single-pixel detector provides a substantially higher temporal sampling rate than 2D arrays \cite{zengHybridGrapheneMetasurfaces2018}, effectively overcoming the temporal bottleneck associated with continuous dynamic imaging. Combined with balanced detection \cite{bacheGhostImagingUsing2004},  the system establishes field correlations computationally by mixing a local oscillator (LO) beam with the spatially integrated bucket signal. Using pre-programmed spatial modulation patterns, this mechanism transposes the spatial interferograms of conventional holography into time-domain coherent detection \cite{ilinaAberrationinsensitiveMicroscopyUsing2019,dengPulsecompressionGhostImaging2016a, longMicrodopplerEffectBased2021}.
	
	Traditional acoustic pressure mapping is inherently intrusive or restricted to surface measurements \cite{sanabriaCalculationVolumetricSound2018, maynard1985nearfield}. While optical techniques exploiting the acousto-optic effect \cite{klein1967unified, torras2012measuring, wang1995continuous} have been explored, achieving rapid and quantitative volumetric reconstructions remains challenging for these methods. Consequently, the visualization of continuous acoustic fields provides a suitable scenario to evaluate the performance of CD-GI, an approach with direct practical applications.
	
	A critical application for this capability is the non-invasive, quantitative visualization of continuous acoustic fields \cite{hasegawaVolumetricAcousticHolography2020, melde2016holograms}. To address this, we experimentally demonstrate in this study the transmissive, quantitative dynamic phase mapping of acoustic fields via field-correlation ghost imaging, enabling deterministic phase calculation through a single-step linear inversion. This approach bypasses the iterative phase-retrieval algorithms required by conventional intensity-detection-based techniques \cite{fienupPhaseRetrievalAlgorithms1982,shechtman2015phase}. By modeling the acoustic field as a dynamic pure-phase object through a discrete matrix-based formulation, we deterministically retrieve acoustic pressure distributions by coupling heterodyne detection and intermediate frequency (IF) spectral analysis \cite{willeminMeasuringAmplitudePhase1983a, rothberg2017international, torras2012measuring} with a pseudo-inverse reconstruction algorithm. The performance of this phase-retrieval-free, single-arm method is systematically evaluated across diverse spatial acoustic modes and validated through comprehensive metrological assessments.
	
	\section{Principle}
	
	\subsection{Subsection title: Single-arm Heterodyne Coherent Detection Computational Ghost Imaging System}
	
	\begin{figure*}[htbp]
		\centering
		\includegraphics[width=\linewidth]{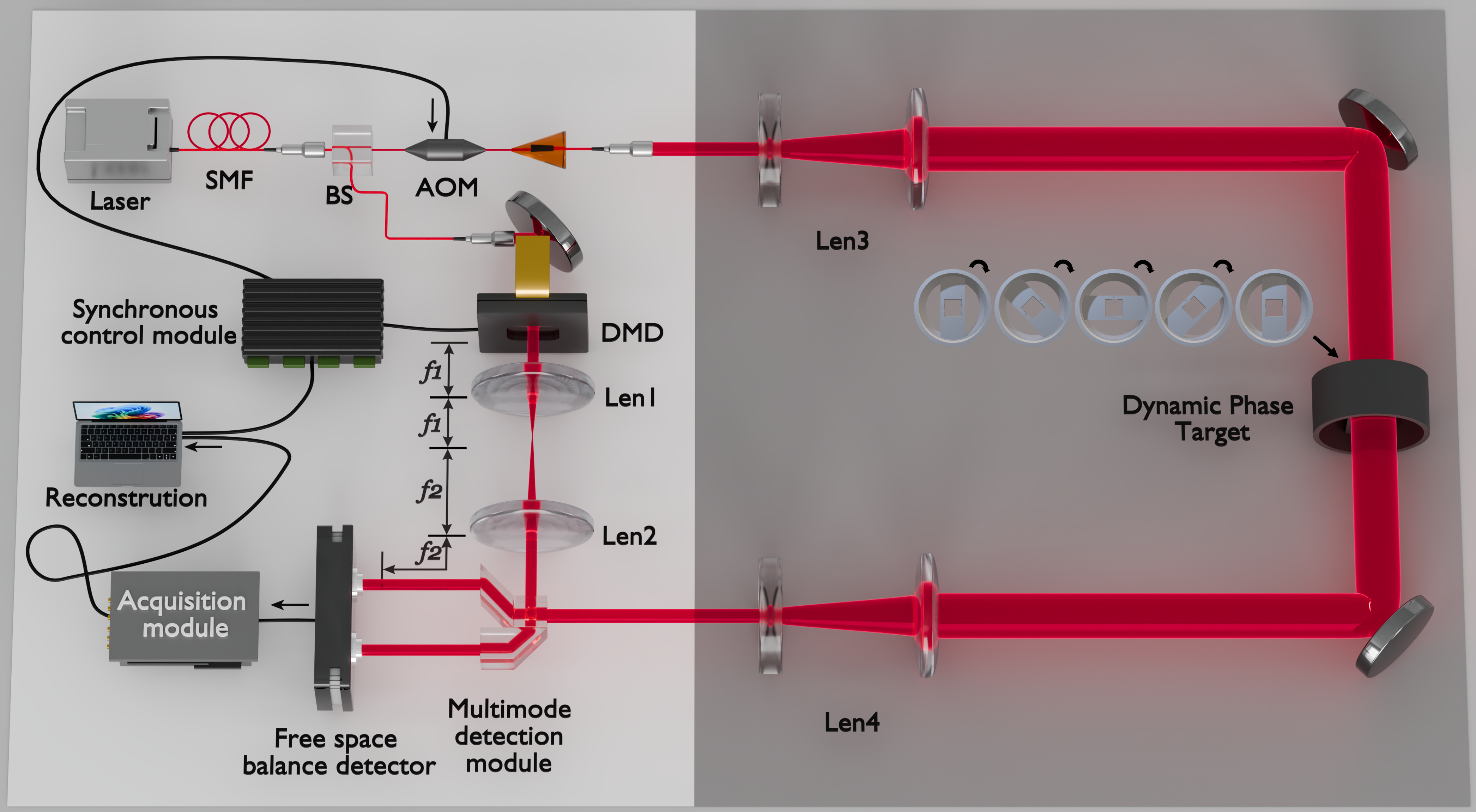} % Please ensure the image extension is supported or omit if not needed during compilation testing
		\caption{Schematic of the experimental CD-GI setup. Laser light, split by a fiber BS, forms the signal and LO paths. In the signal path, AOM-shifted light probes the programmable ultrasonic dynamic phase target, with scattered light collected via an imaging system. In the modulation path, the LO is spatially encoded by a DMD, relayed by a 4f system, and combined with the signal at a BPD for heterodyne detection. Synchronous control ensures phase stability. (Abbreviations: SMF, single-mode fiber; BS, beam splitter; AOM, acousto-optic modulator; FL, focal length; DMD, digital micromirror device; M, mirror; BPD, balanced photodetector; OA, optical amplifier; Len, lens)}
		\label{fig:setup}
	\end{figure*}
	
	The proposed single-arm CD-GI system employs a post-modulation heterodyne architecture to achieve quantitative dynamic phase mapping. The overall optical layout and signal flow are schematically illustrated in Fig.~\ref{fig:setup}. In this single-arm architecture, only the object path traverses the target environment, while the LO beam is routed locally at the receiver.
	
	The experimental setup centers around a narrow-linewidth laser operating at 1550 nm. The coherent laser output is delivered via a single-mode fiber (SMF) and split into an illumination path, which constitutes the object arm for CD-GI, and an LO path using a fiber beam splitter (BS), leveraging robust optical fiber sensor technology \cite{giallorenzi1982optical} to establish a strict phase-coherent relationship between the two beams.
	
	In the object path, the illumination beam first passes through an acousto-optic modulator (AOM), which introduces a stable frequency shift $f_{\mathrm{if}} = 2\ \mathrm{MHz}$ to the optical carrier, setting the IF for heterodyne detection. The frequency-shifted beam is subsequently amplified by an optical amplifier and free-space collimated to probe the acoustic volume of interest. Upon interacting with the dynamically varying phase of the target, the beam undergoes a complex phase modulation. We denote the continuous optical field transmitted directly after the dynamic phase target by its complex transmission function $O(t, \boldsymbol{\rho})$, where $t$ and $\boldsymbol{\rho}$ represent time and the two-dimensional spatial coordinates, respectively. Following this interaction, the phase-modulated object field is collected and guided by a dedicated receiving imaging system. As shown in the setup, this system consists of reflection mirrors and a series of lenses—specifically, Len3 and associated optics—designed to image the target plane onto the detection plane of a free-space balanced photodetector. Owing to the finite aperture of these receiving optics, the optical propagation inherently acts as a low-pass spatial filter governed by the system's coherent transfer function (CTF). The associated coherent amplitude point spread function (PSF), denoted as $h(\boldsymbol{\rho})$, serves as the spatial convolution kernel. For a circular limiting aperture, this PSF manifests as an Airy disk pattern, proportional to $J_1(k \mathrm{NA} |\boldsymbol{\rho}|) / (k \mathrm{NA} |\boldsymbol{\rho}|)$, where $J_1$ is the first-order Bessel function of the first kind, $k$ is the wavenumber, and $\mathrm{NA}$ is the numerical aperture \cite{zhengWidefieldHighresolutionFourier2013}. Consequently, the complex amplitude of the signal field at the detector plane is a diffraction-limited representation of the target, mathematically defined by the continuous 2D spatial convolution $\widetilde{O}(t, \boldsymbol{\rho}_{\mathrm{d}}) = (h * O)(t, \boldsymbol{\rho}_{\mathrm{d}})$. The complete spatio-temporal signal field at the detector plane is thus:
	\begin{equation}
		E_{\mathrm{d}}(t, \boldsymbol{\rho}_{\mathrm{d}}) = \widetilde{O}(t, \boldsymbol{\rho}_{\mathrm{d}}) \exp[\mathrm{i} 2\pi (f_{\mathrm{c}} + f_{\mathrm{if}}) t].
	\end{equation}
	
	Contrary to traditional pre-modulation schemes where spatial encoding is applied prior to target interaction, our configuration implements spatial encoding directly on the LO beam to form the post-modulation architecture. The split LO beam of GI is collimated and uniformly illuminates a digital micromirror device (DMD), which is programmed to display predetermined sequences of complex spatial basis patterns by switching its micro-mirrors. Subsequently, this spatially encoded LO field is relayed from the DMD plane to the photosensitive surface of the free-space balanced photodetector via a precise 4f imaging system, comprising lenses Len1 and Len2. Separated by $f_1 + f_2$, this system ensures a distortion-free image of the DMD pattern is formed on the detector. While a wide-open aperture at the Fourier plane minimizes spatial filtering of the LO, a specifically tailored aperture could be introduced to modify the spatial frequencies of the encoding patterns. In our implementation, we design and maximize the transfer of spatial information from the LO path, relying primarily on the intrinsic CTF of the signal path to define the fundamental resolution limit. The spatially encoded LO field on the detector plane for the modulation pattern $\tau$ can be expressed as:
	\begin{equation}
		E_{\mathrm{LO}}(t, \boldsymbol{\rho}_{\mathrm{d}}; \tau) = \sqrt{\eta_{\mathrm{LO}}} A_{0} P(\boldsymbol{\rho}_{\mathrm{d}}; \tau) \exp(\mathrm{i} 2\pi f_{\mathrm{c}} t),
	\end{equation}
	where $\eta_{\mathrm{LO}}$ is the LO power ratio, $A_{0}$ is the initial laser amplitude, and $P(\boldsymbol{\rho}_{\mathrm{d}}; \tau)$ represents the spatial encoding pattern $\tau$ projected onto the detection plane coordinates $\boldsymbol{\rho}_{\mathrm{d}}$. Functionally, by operating at the receiver end, this architecture isolates the spatial modulator from an illumination beam with high power. This directly bypasses the optical damage threshold limits present in traditional schemes that modulate light at the transmitter. It also transforms a local oscillator with a single spatial mode into a structured one with multiple modes that selectively couple with the spatial profile of the scattered light, enhancing the effective etendue of the coherent receiver.
	
	Within the free-space balanced photodetector module, the band-limited signal field $E_{\mathrm{d}}(t, \boldsymbol{\rho}_{\mathrm{d}})$ and the spatially encoded LO field $E_{\mathrm{LO}}(t, \boldsymbol{\rho}_{\mathrm{d}}; \tau)$ are spatially mixed to generate an interference pattern. The dual-path differential output of the balanced detector cancels the large DC background components and isolates the coherent beat frequency. The resulting photocurrent $i(t; \tau)$ for the mask $\tau$ is modeled as the spatial integration of the complex interference terms over the detector's finite active area $A_{\mathrm{d}}$ \cite{qiMultimodeCoherentDetection2026a}:
	\begin{align}
		i(t; \tau) &= 2 R_{L} \sqrt{\eta_{\mathrm{LO}}} A_{0} \notag\\ &\times \mathrm{Re} \left\{ \exp(-\mathrm{i} 2\pi f_{\mathrm{if}} t) \int_{A_{\mathrm{d}}} \widetilde{O}^{*}(t, \boldsymbol{\rho}_{\mathrm{d}}) P(\boldsymbol{\rho}_{\mathrm{d}}; \tau) \mathrm{d}^2\boldsymbol{\rho}_{\mathrm{d}} \right\},
	\end{align}
	where $R_{L}$ is the detector's responsivity and conversion factor, and $\mathrm{Re}\{\cdot\}$ denotes the real part. This resulting IF signal is digitized by an acquisition module, with phase stability maintained by a synchronous control module coordinating data acquisition, DMD switching, and the AOM driver.
	
	By performing a Fourier transform on this time-domain photocurrent and extracting the component at $f_{\mathrm{if}}$, we obtain the complex phasor of the IF signal. To transition into a computational imaging framework, the continuous integrated fields are discretized into a matrix representation \cite{edgarPrinciplesProspectsSinglepixel2019}. Let the continuous filtered object field $\widetilde{O}(t, \boldsymbol{\rho}_{\mathrm{d}})$ be mapped to a discrete matrix $\widetilde{\boldsymbol{O}}(t) \in \mathbb{C}^{N_x \times N_y}$, and the spatial pattern $P(\boldsymbol{\rho}_{\mathrm{d}}; \tau)$ be mapped to a modulation matrix $\boldsymbol{P}(\tau) \in \mathbb{C}^{N_x \times N_y}$. Accounting for systemic complex Gaussian noise $n(t; \tau)$, and leveraging the matrix trace to represent the spatial inner product, the complex coherent output signal from a single bucket detector, denoted as $y(t; \tau)$, is formulated as \cite{wang2012coherent}:
	\begin{equation}
		y(t; \tau) = \mathrm{tr}(\boldsymbol{P}^{\mathrm{H}}(\tau) \widetilde{\boldsymbol{O}}(t)) + n(t; \tau),
	\end{equation}
	where $\boldsymbol{P}^{\mathrm{H}}(\tau)$ is the conjugate transpose of the modulation matrix $\boldsymbol{P}(\tau)$. This model translates the physical coherent integration into a standard linear inverse problem framework.
	
	\subsection{Acousto-Optic Diagnostics and Spectral Reconstruction}
	
	Building upon this mathematical model, we explicitly specify the target matrix $\boldsymbol{O}(t)$ for the current application. The target is a dynamic, distributed phase modulation induced by the acoustic field via the acousto-optic effect.
	
	The acoustic pressure field $p(t, \boldsymbol{\rho})$ at position $\boldsymbol{\rho}$ and time $t$ is expressed as a monochromatic wave:
	$p(t, \boldsymbol{\rho}) = A_p(\boldsymbol{\rho})\cos(\omega_{a}t)$, where $A_p(\boldsymbol{\rho})$ represents the spatial pressure amplitude distribution, and $\omega_a$ denotes the angular frequency of the acoustic wave. This pressure field modulates the refractive index $n$ of the medium according to $\Delta n(t, \boldsymbol{\rho}) = \eta \cdot p(t, \boldsymbol{\rho})$, with $\eta$ being the acousto-optic coefficient \cite{klein1967unified, resink2012state}. A coherent optical probe beam (vacuum wavenumber $k_0 = 2\pi/\lambda_0$) traversing this modulated region of length $L$ accumulates a spatially and temporally varying phase shift:
	\begin{equation}
		\Delta \phi(t, \boldsymbol{\rho}) \approx k_0 L \eta A_p(\boldsymbol{\rho})\cos(\omega_{a}t).
	\end{equation}
	By discretizing the spatial coordinates into a matrix format, we define the acoustic phase-modulation amplitude matrix as $\boldsymbol{A} = k_0 L \eta \boldsymbol{A}_p$. The complex transmission function of the acoustic field acts as a pure phase object within the weak-scattering approximation \cite{devaney1981inverse}, represented by the dynamic matrix: $\boldsymbol{O}(t) = \exp\left[\mathrm{i}\boldsymbol{A}\cos(\omega_{a}t)\right]$. This dynamic phase object is illuminated with a sequence of structured patterns generated by the DMD. Let $\boldsymbol{P}(\tau)$ denote the $\tau$-th binary illumination matrix. Since the DMD performs real-valued amplitude modulation, the conjugate transpose simplifies to the regular transpose ($\boldsymbol{P}^{\mathrm{H}}(\tau) = \boldsymbol{P}^{\mathrm{T}}(\tau)$).
	
	Utilizing the trace-based inner product formulation, the resulting IF bucket signal $y_{\mathrm{if}}(t; \tau)$ is expressed as the trace of the matrix product between the transposed illumination pattern and the dynamic phase object:
	
	\begin{equation}
		y_{\mathrm{if}}(t; \tau) \propto e^{\mathrm{i}\varphi} \, \mathrm{tr}\left\{ \boldsymbol{P}^{\mathrm{T}}(\tau) \exp\left[\mathrm{i}\omega_{\mathrm{if}}t + \mathrm{i}\boldsymbol{A}\cos(\omega_{a}t)\right] \right\},
	\end{equation}
	where $\varphi$ is a constant, system-dependent phase offset.
	
	With the acoustic modulation successfully encoded into the time-domain heterodyne signal $y_{\mathrm{if}}(t; \tau)$, we next extract the embedded spatial acoustic phase matrix $\boldsymbol{A}$ by analyzing the signal in the frequency domain. Applying the Jacobi-Anger expansion \cite{willeminMeasuringAmplitudePhase1983a,torras2012measuring}, $\exp(\mathrm{i}z\cos\theta) = \sum_{m=-\infty}^{\infty} \mathrm{i}^m J_m(z) e^{\mathrm{i} m\theta}$, element-wise to the dynamic phase matrix $\boldsymbol{O}(t)$, the spectrum of the continuous-time IF signal is derived as:
	
	\begin{widetext}
		\begin{align}
			\mathcal{F}_{t}\{y_{\mathrm{if}}(t;\tau)\} &\propto e^{\mathrm{i}\varphi} \, \mathrm{tr}\left\{ \boldsymbol{P}^{\mathrm{T}}(\tau) \mathcal{F}_{t} \bigg\{ \exp\left[\mathrm{i}\omega_{\mathrm{if}}t + \mathrm{i}\boldsymbol{A}\cos(\omega_{a}t)\right] \bigg\} \right\} \notag\\
			&= e^{\mathrm{i}\varphi} \, \mathrm{tr}\left\{ \boldsymbol{P}^{\mathrm{T}}(\tau) \mathcal{F}_{t} \bigg\{ \sum_{m=-\infty}^{\infty} \mathrm{i}^{m} J_{m}(\boldsymbol{A}) \exp\left[\mathrm{i}\left(\omega_{\mathrm{if}} + m\omega_{a}\right)t\right] \bigg\} \right\} \notag\\
			&= 2\pi e^{\mathrm{i}\varphi} \sum_{m=-\infty}^{\infty} \mathrm{i}^{m} \, \mathrm{tr}\left\{ \boldsymbol{P}^{\mathrm{T}}(\tau) J_{m}(\boldsymbol{A}) \right\} \delta\left( \omega - \left( \omega_{\mathrm{if}} + m \omega_{a} \right) \right),
		\end{align}
	\end{widetext}
	
	where $J_m(\boldsymbol{A})$ denotes the element-wise application of the $m$-th order Bessel function of the first kind to the acoustic modulation matrix.
	
	The resulting spectrum features discrete spectral lines located at frequencies $\omega = \omega_{\mathrm{if}} + m\omega_a$. Analysis focuses on the central carrier component ($m=0$) and the two first-order sidebands ($m=\pm1$), as they encode the primary physical information of the acoustic modulation. In the discrete experimental setting, let the sampling frequency be $f_s$, yielding a discrete Fourier transform (DFT) frequency resolution of $\Delta \omega = 2\pi f_s / N$, where $N$ is the total number of samples. The frequency bin indices corresponding to the intermediate and acoustic frequencies are $k_{\mathrm{if}} = \mathrm{round}(\omega_{\mathrm{if}}/\Delta \omega)$ and $k_a = \mathrm{round}(\omega_a/\Delta \omega)$, respectively.
	
	By computing the DFT, denoted as $Y[\varepsilon] = \sum_{n=0}^{N-1} y_{\mathrm{if}}[n] \exp(-\mathrm{i} 2\pi \varepsilon n / N)$, the complex amplitudes corresponding to the three spectral peaks of interest are extracted at bins $k_{-1} = k_{\mathrm{if}} - k_a$, $k_{0} = k_{\mathrm{if}}$, and $k_{+1} = k_{\mathrm{if}} + k_a$ \cite{longMicrodopplerEffectBased2021, torras2012measuring}:

		\begin{align}
			Y_{-1}(\tau) &= Y[k_{-1}] \approx -\alpha \mathrm{i} \cdot \mathrm{tr}\left\{ \boldsymbol{P}^{\mathrm{T}}(\tau) J_{-1}(\boldsymbol{A}) \right\}, \notag\\
			Y_{0}(\tau)  &= Y[k_{0}]  \approx \alpha \cdot  \mathrm{tr}\left\{ \boldsymbol{P}^{\mathrm{T}}(\tau) J_{0}(\boldsymbol{A}) \right\}, \notag\\
			Y_{+1}(\tau) &= Y[k_{+1}] \approx  \alpha \mathrm{i} \cdot \mathrm{tr}\left\{ \boldsymbol{P}^{\mathrm{T}}(\tau) J_{1}(\boldsymbol{A}) \right\},  \label{phase}
		\end{align}

	where $\alpha$ is a complex scaling factor accounting for discretization, windowing, DFT normalization, and the system's initial phase offset $\varphi$.
	
	The recovery of the spatial matrix $\boldsymbol{A}$ from a sequence of $M$ spectral measurements $\{Y_m(\tau)\}_{\tau=1}^M$ is formulated as a standard linear inverse problem. To establish this mathematical framework, we vectorize the 2D spatial matrices into 1D column vectors. Let $\mathbf{y}_m = \frac{1}{\alpha \mathrm{i}^{m}} [Y_m(1), Y_m(2), \dots, Y_m(M)]^{\mathrm{T}} \in \mathbb{C}^{M \times 1}$ denote the measurement vector for the $m$-th IF component. We then construct the sensing matrix $\mathbf{H} \in \mathbb{R}^{M \times (N_x N_y)}$, where each row corresponds to the transposed vectorized modulation pattern, $\mathrm{vec}(\boldsymbol{P}(\tau))^{\mathrm{T}}$.
	
	Accounting for the additive measurement noise $\mathbf{n}_m \in \mathbb{C}^{M \times 1}$ and absorbing scalar constants into $\mathbf{H}$, the forward measurement process and the subsequent spatial reconstruction using the Moore-Penrose pseudo-inverse algorithm \cite{katz2009compressive,duarteSinglepixelImagingCompressive2008,tongPreconditionedDeconvolutionMethod2021,fesslerSpacealternatingGeneralizedExpectationmaximization1994} are jointly expressed as:
	
	\begin{align} 
		\mathbf{y}_m &= \mathbf{H} \mathrm{vec}(J_m(\boldsymbol{A})) + \mathbf{n}_m, \\ \mathrm{vec}(\widehat{J_m(\boldsymbol{A})}) &= \mathbf{H}^+ \mathbf{y}_m,
	\end{align}
	where $\mathbf{H}^+ = (\mathbf{H}^{\mathrm{T}}\mathbf{H})^{-1}\mathbf{H}^{\mathrm{T}}$ represents the pseudo-inverse of the sensing matrix. After reshaping the estimated 1D vector $\mathrm{vec}(\widehat{J_m(\boldsymbol{A})})$ back into the 2D spatial matrix $\widehat{J_m(\boldsymbol{A})}$, the acoustic phase-modulation amplitude $\boldsymbol{A}$, which maps the quantitative phase distribution, could be calculated by analytically inverting the corresponding Bessel function. Assuming weak acoustic modulation ($\Vert\boldsymbol{A}\Vert_{\infty} \ll 1$), this inversion is linearized via small-argument approximations \cite{monchalin1986optical}:
	\begin{equation}
		J_0(\boldsymbol{A}) \approx \mathbf{1} - \boldsymbol{A}^{\circ 2}/4 ,\quad J_1(\boldsymbol{A})=-J_{-1}(\boldsymbol{A}) \approx \boldsymbol{A}/2 , \label{app}
	\end{equation}
	where $\circ$ denotes the Hadamard power and $\mathbf{1}$ is the all-ones matrix. While $\boldsymbol{A}$ can theoretically be recovered from any of these spectral components, the reconstruction derived from the IF carrier ($m=0$) is often degraded by system noise and phase fluctuations, as the minute acoustic term $-\boldsymbol{A}^{\circ 2}/4$ is superimposed on a large DC background $\mathbf{1}$. Conversely, the sideband signals ($Y_{-1}$ and $Y_{+1}$) are directly proportional to $\boldsymbol{A}/2$ against a near-zero background. Therefore, performing the pseudo-inverse reconstruction exclusively on the sideband signals provides a fundamentally higher signal-to-noise ratio for mapping the quantitative dynamic phase distribution.
	
	\section{Results}
	
	\subsection{Direct Spatial Dynamic Phase Mapping of Acoustic Fields}
	
	To evaluate the spatial dynamic phase mapping capability of the proposed CD-GI architecture and validate the theoretical spectral reconstruction model, we interrogated a complex acoustic geometry generated by a programmable acoustic levitator. Our heterodyne coherent detection scheme directly extracts the complex optical field from the IF beat signal, enabling deterministic, quantitative dynamic phase mapping without relying on iterative phase retrieval algorithms.
	
\begin{figure*}[htbp]
		\centering
		\includegraphics[width=\textwidth]{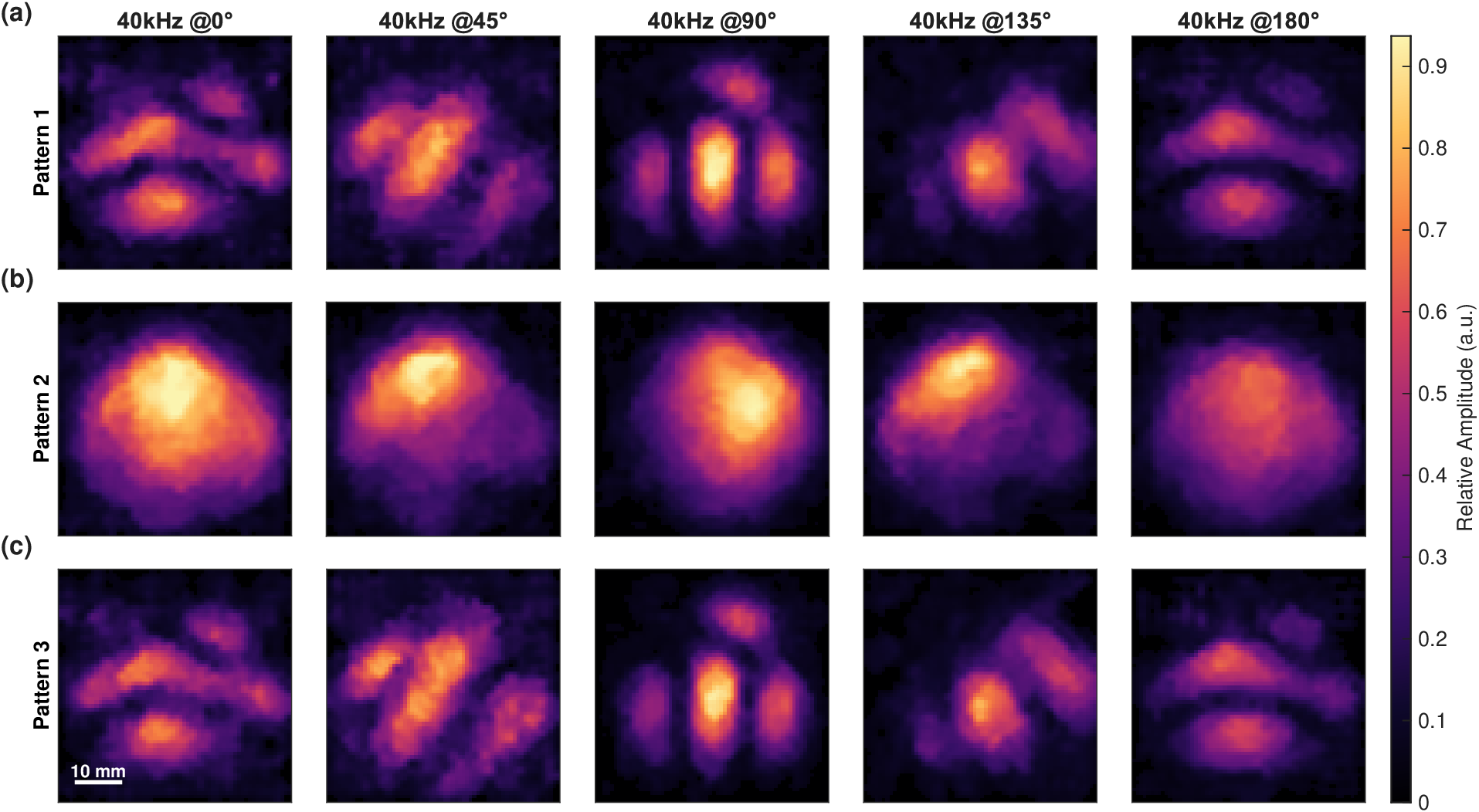}
		\caption{Experimental reconstructions of the acoustic pressure distributions at 40 kHz, demonstrating the imaging fidelity across different IF spectral components. (a)--(c) Reconstructions derived from the lower heterodyne sideband $Y_{-1}$ (Row 1), the central carrier $Y_0$ (Row 2), and the upper sideband $Y_{+1}$ (Row 3), respectively. Each row presents the retrieved acoustic field at five in-plane projection angles ($0^{\circ}$, $45^{\circ}$, $90^{\circ}$, $135^{\circ}$, and $180^{\circ}$). Consistent with the theoretical model, the sideband reconstructions (Rows 1 and 3) exhibit significantly higher signal-to-noise ratios, whereas the carrier reconstruction (Row 2) is severely degraded by the constant DC background. The Magma colormap indicates the relative acoustic pressure amplitude. The white scale bar denotes a spatial length of 10 mm.}
		\label{fig:modes}
	\end{figure*}

	Driven by a square array of ultrasonic transducers with a continuously tunable frequency, this levitator configuration generates highly structured, multidirectional acoustic standing waves with intricate spatial symmetries, providing a stringent testbed for evaluating imaging fidelity. Operating the array at a representative frequency of 40 kHz, the CD-GI system mapped the horizontal cross-sectional plane within the levitation chamber. By utilizing the temporal acquisition of the coherent receiver, the IF signal was continuously digitized at a sampling rate of $125~\mathrm{MS/s}$. This temporal mapping allowed the system to demodulate and freeze the steady-state spatial amplitude profiles of the high-frequency acoustic pressure fields, downconverting the $40~\mathrm{kHz}$ microsecond-scale optical phase oscillations ($\sim 25~\mu\mathrm{s}$ period) into stationary sideband components. As shown in Fig.~\ref{fig:modes}, the acquired time-domain signals were processed to independently reconstruct the acoustic field from the three IF spectral components ($Y_{-1}$, $Y_0$, and $Y_{+1}$). Based on the linear Bessel approximation (see Eq.~\eqref{app}), the mapped relative amplitude is directly proportional to the dynamic optical phase shift ($\Delta\phi$).
	
	 Consistent with the theoretical framework, the pseudo-inverse reconstructions derived from the first-order sidebands (Rows 1 and 3) yielded maps of the dynamic acoustic pressure field with high fidelity and a high ratio of signal to noise across all projection angles ($0^\circ$ to $180^\circ$). In contrast, the reconstruction derived from the central carrier (Row 2) suffered from severe image degradation. This degradation arises from noise at low frequencies near the central carrier, dominated by environmental thermal drifts, ambient vibrations, and the relative intensity noise of the laser. Heterodyne extraction at the first-order sidebands circumvents these baseband disturbances, isolating the linear acoustic term against a vanishing background to preserve quantitative fidelity \cite{monchalin1986optical}.
	
	Next, to quantitatively assess the frequency-dependent spatial resolution, the transducer array of the acoustic levitator was configured to generate continuous planar acoustic fields at discrete driving frequencies (25 kHz, 31.25 kHz, and 40 kHz). Corresponding numerical models were established using the commercial finite element analysis software COMSOL Multiphysics~\cite{COMSOLMultiphysics622023}, applying boundary conditions consistent with the experimental acoustic environment.
	
\begin{figure}[htbp]
		\centering
		\includegraphics[width=\linewidth]{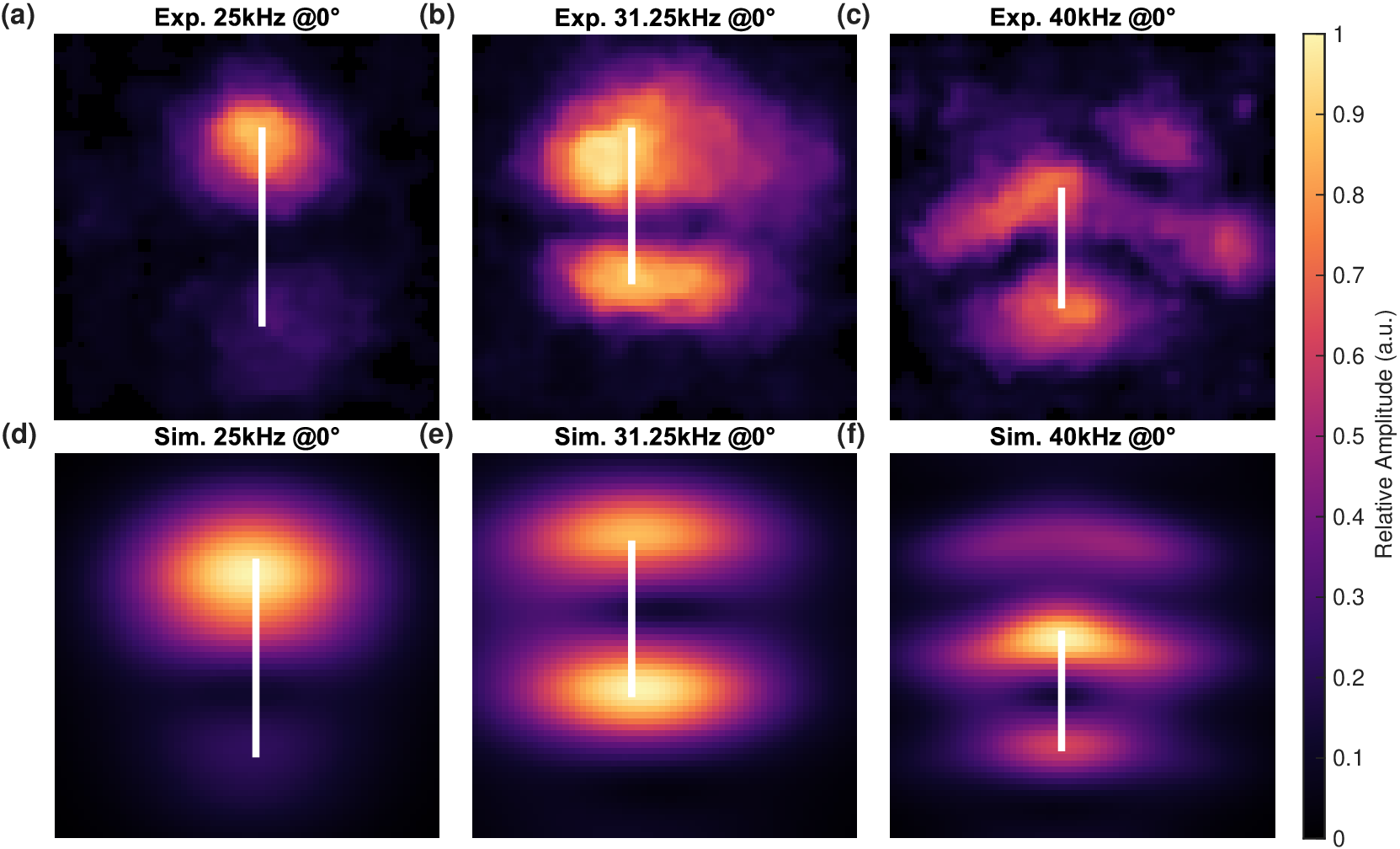}
		\caption{Comparison of experimental and simulated acoustic heatmaps at a $0^{\circ}$ projection angle. (a)--(c) CD-GI experimental reconstructions of the acoustic pressure field at 25 kHz, 31.25 kHz, and 40 kHz, respectively. (d)--(f) Corresponding numerical simulations performed using COMSOL Multiphysics \cite{COMSOLMultiphysics622023}. The vertical white bars across all panels indicate an identical spatial length, serving as a uniform scale reference to illustrate the frequency-dependent spatial periodicity. The high spatial correlation validates the imaging fidelity across the operational frequency bandwidth.}
		\label{fig:heatmaps}
	\end{figure}

	For each frequency configuration, the CD-GI system executed a full-field interrogation. Applying the optimized sideband pseudo-inverse reconstruction, the projected phase measurements were synthesized into two-dimensional acoustic pressure cross-sections. As depicted in Fig.~\ref{fig:heatmaps}, the experimental reconstructions exhibit excellent structural correspondence with the numerical simulations. The optical system accurately resolved the periodic fringes of the dynamic phase distribution, clearly illustrating that modifications to the driving frequency proportionally alter the spatial periodicity of the field pattern, all directly achieved through a single-arm architecture without computational phase retrieval algorithms.
	
	\subsection{Quantitative Evaluation of the Reconstructions}
	
	\begin{figure*}[htbp]
		\centering
		\includegraphics[width=\textwidth]{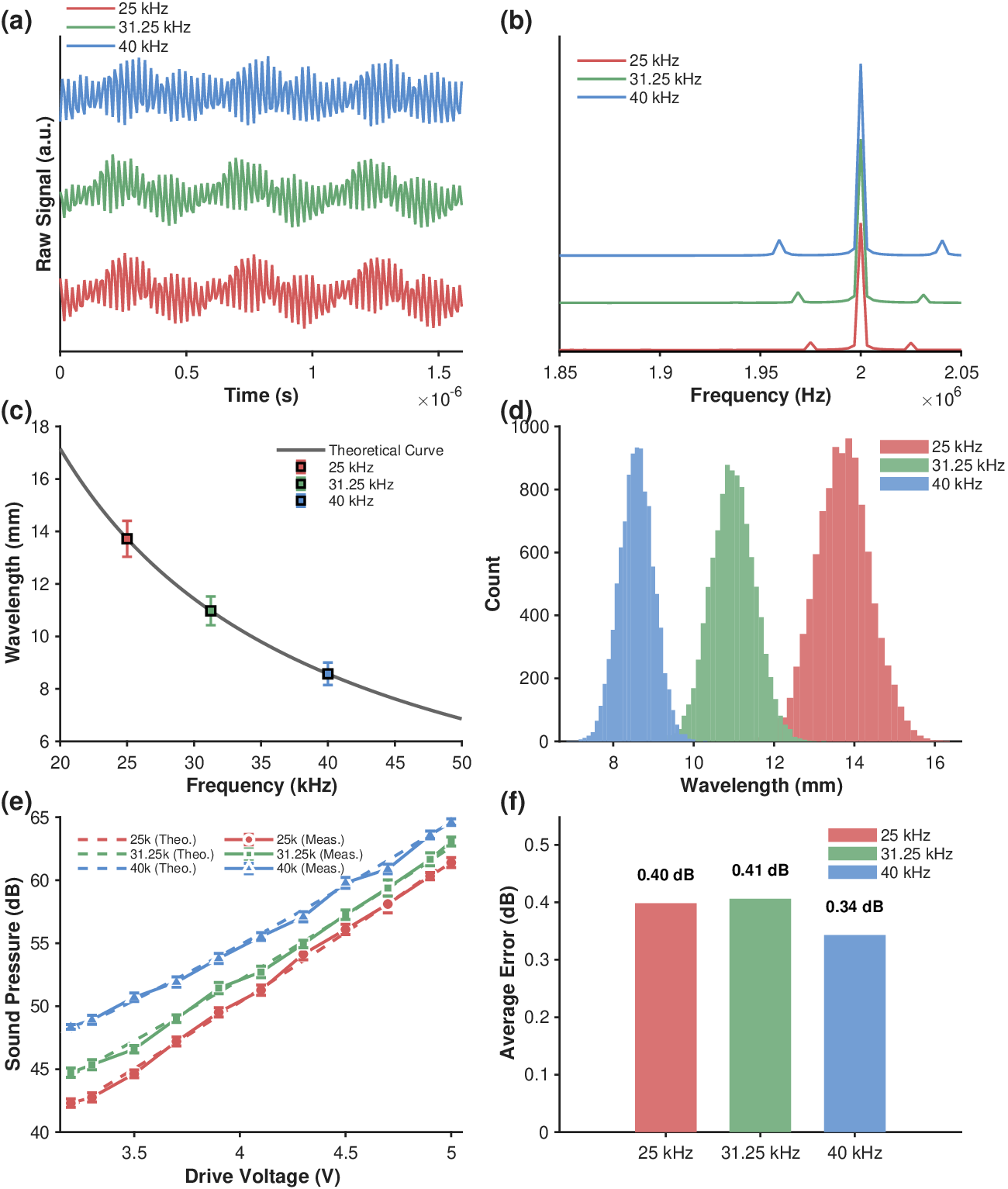}
		\caption{Quantitative phase analysis of the CD-GI acoustic field reconstructions. (a) Raw time-domain heterodyne signals captured under a constant IF for acoustic driving frequencies of 25~kHz, 31.25~kHz, and 40~kHz. (b) DFT spectra of the detected heterodyne signals, averaged over 100 repetitions. (c) Measured acoustic wavelength as a function of driving frequency compared against the theoretical dispersion curve, based on 100 repeated experiments. (d) Statistical distribution (histogram) of the morphological wavelengths extracted from 150 time-window segments across 100 repeated experiments. (e) Measured sound pressure level (SPL) response versus drive voltage averaged across 100 repetitions, compared with theoretical curves. (f) Average measurement error (dB) for each targeted frequency, quantified over 100 independent experimental repetitions.}
		\label{fig:analysis}
	\end{figure*}

	Note that the spatial distributions presented in Figs.~\ref{fig:modes} and \ref{fig:heatmaps} characterize the relative magnitude of the acoustic pressure field. Governed by the acousto-optic effect and the linear Bessel approximation, these reconstructed profiles are directly proportional to the dynamic optical phase shift ($\Delta\phi$) accumulated by the probing beam. Typically, translating this relative mapping to absolute phase requires a calibration in situ to determine the proportionality constant. While this constant can be theoretically estimated by extracting the effective acousto-optic interaction length via multiphysics simulations (e.g., COMSOL Multiphysics \cite{COMSOLMultiphysics622023}), directly linking the recovered signal to the physical phase. We report the results in relative units (a.u.) to allow direct comparison with acoustic simulations.
	
	To quantitatively evaluate the reconstruction accuracy, the acoustic fields generated at 25 kHz, 31.25 kHz, and 40 kHz were analyzed, as summarized in Fig.~\ref{fig:analysis}. The primary parameters evaluated were the heterodyne spectral fidelity, the optically extracted acoustic wavelength, and the relative sound pressure level (SPL), defined as $\mathrm{SPL} = 20\log_{10}(A_p / A_{p,\mathrm{ref}})$, where $A_{p,\mathrm{ref}}$ is the reference pressure amplitude.
	
	Figure~\ref{fig:analysis}(a) presents the raw time-domain heterodyne signals captured under a constant IF for acoustic driving frequencies of 25~kHz, 31.25~kHz, and 40~kHz. Their corresponding DFT spectra, averaged over 100 repetitions, are displayed in Fig.~\ref{fig:analysis}(b), where the targeted acoustic driving frequencies are resolved with a high SNR. This spectral fidelity stems from the inherently high time-bandwidth product of the coherent detection scheme, which acts as a temporal pulse compressor to efficiently isolate acousto-optic phase modulations from ambient environmental noise. This mechanism enhances phase sensitivity, enabling the deterministic resolution of subtle optical dynamic phase perturbations induced by low-pressure acoustic waves. The spatial acoustic wavelength was subsequently extracted from the spatial periodicity of the reconstructed fringe patterns. As shown in Fig.~\ref{fig:analysis}(c), the optically measured wavelengths, based on 100 repeated experiments, agree well with the theoretical dispersion curve calculated using the ambient speed of sound in air. Additionally, the statistical distribution in Fig.~\ref{fig:analysis}(d) details the morphological wavelengths extracted from 150 time-window segments across these 100 repeated experiments, verifying the system's robust spatial resolving capabilities.
	
	Furthermore, the SPL was quantitatively calibrated directly from the measured optical phase shift. By sweeping the array's drive voltage, a linear correlation was established between the drive voltage and the optically reconstructed sound pressure, as depicted in Fig.~\ref{fig:analysis}(e), which compares the measured response averaged across 100 repetitions against theoretical curves. Finally, based on this comprehensive set of 100 independent measurements, the average measurement deviation between the theoretical predictions and experimental reconstructions is quantified in Fig.~\ref{fig:analysis}(f). The observed deviations confirm that the CD-GI system maintains quantitative accuracy across the tested acoustic intensities and frequencies. By directly extracting the spatial sideband amplitude, the framework avoids $2\pi$ phase-wrapping artifacts and exhibits robustness against optical aberrations \cite{ilinaAberrationinsensitiveMicroscopyUsing2019}. As a result, the primary nonlinear limit of the system arises from the Bessel expansion. The linear small-argument approximation ($J_1(\boldsymbol{A}) \approx \boldsymbol{A}/2$) restricts the maximum phase shift to $\Vert\boldsymbol{A}\Vert_\infty \lesssim 0.63$ rad to maintain truncation errors below $5\%$. Relaxing this linear constraint (i.e., applying the inverse Bessel function $J_1^{-1}$ after the matrix inversion) extends the monotonic measurement limit to $1.84$ rad, corresponding to the first stationary point of $J_1$.

	\section{Discussion}
	
  The current CD-GI architecture successfully demonstrates high-fidelity quantitative mapping of dynamic phase distributions. To fully contextualize its capabilities, we characterize the physical parameters governing its operation and identify key avenues for performance scaling.
  
  Spatially, the system's resolution depends on the stricter of two limits: the optical CTF ($f_{\mathrm{opt}} = \mathrm{NA}/\lambda \approx 129~\mathrm{lp/mm}$ for the $\mathrm{NA}=0.2$ receiving optics) and the computational Nyquist limit determined by the DMD's spatial sampling rate at the conjugate plane ($f_{\mathrm{DMD}} = 1/(2\Delta x_{\mathrm{eff}}) \approx 6.9~\mathrm{lp/mm}$) \cite{tongPreconditionedDeconvolutionMethod2021}. Due to the trade-off between the physical field of view (FOV) and spatial sampling density, the system operates optimally when the FOV covers 2 to 10 resolvable spatial periods of the target phase field.
  
  Temporally, although the raw intermediate-frequency (IF) beat signal is continuously digitized at $125~\mathrm{MS/s}$, the effective 2D imaging frame rate is currently restricted by the DMD's mechanical switching limit ($T_{\mathrm{DMD}} \approx 45~\mu\mathrm{s}$) and the necessary heterodyne signal integration time \cite{liCoprimefrequenciedSinusoidalModulation2016}. Our framework remains effective for non-stationary dynamic phases, such as complex evolutions that deviate from a strict single-harmonic model. Owing to the mathematical completeness of the Bessel series \cite{cohenTimeFrequencyAnalysis1995}, these non-stationary processes can be linearly expanded and approximated within a brief observation window. Therefore, provided the total acquisition time for a 2D frame ($ M \cdot \max(T_{\mathrm{DMD}}, T_{\mathrm{window}})$) is shorter than the characteristic evolution time of the phase, the system can resolve these transient dynamics by isolating and synthesizing multiple frequency-domain channels.
  
  Beyond hardware specifications, the system's performance is underpinned by its computational stability. The spatial pseudo-inverse ($\mathbf{H}^+$) remains robust because the pure binary $\{0, 1\}$ DMD projection ensures the sensing matrix $\mathbf{H}$ rigorously acts as a random matrix drawn from the Bernoulli ensemble. According to non-asymptotic random matrix theory, such sub-Gaussian matrices possess favorable singular value spectra, with the smallest singular value stochastically bounded away from zero \cite{vershynin2018high, tao2010random}. This characteristic naturally bounds the condition number $\kappa(\mathbf{H})$, effectively suppressing the noise amplification that typically plagues ill-posed inverse problems. Complementing this mathematical stability, the single-pixel architecture inherently provides physical resilience against environmental distortions. While localized phase perturbations introduce spatially random phase errors across the coherent detector array plane, thereby degrading the detection accuracy of individual pixels, our bucket detector spatially integrates the field. Consequently, by the central limit theorem, independent spatial phase fluctuations (e.g., turbulence or random aberrations \cite{ilinaAberrationinsensitiveMicroscopyUsing2019,edgarPrinciplesProspectsSinglepixel2019}) statistically average out, suppressing uncorrelated environmental noise.
  
  Building on this mathematical foundation, the optical architecture provides clear pathways for performance scaling. Algorithmically, integrating compressive sensing (CS) strategies \cite{duarteSinglepixelImagingCompressive2008, katz2009compressive} can reduce acquisition times and surpass traditional limits to achieve super-resolution \cite{chenMulticolorSuperresolutionStructured2025, tsangQuantumTheorySuperresolution2016}, while temporal windowing algorithms can extract finer evolutionary details without hardware alterations. On the hardware front, replacing conventional DMDs with fast metasurface modulators \cite{zengHybridGrapheneMetasurfaces2018} can eliminate mechanical bottlenecks, facilitating instantaneous phase mapping of transient events like shockwaves \cite{hargather2012comparison, liangSingleshotStereopolarimetricCompressed2020}. Finally, by scaling the NA of the relay optics, this framework transitions to field-correlation microscopy \cite{ilinaAberrationinsensitiveMicroscopyUsing2019,chenMulticolorSuperresolutionStructured2025}. This versatility enables the CD-GI platform to probe non-harmonic, micro-scale dynamics across diverse transparent media, including complex biological and microfluidic systems.
	
	\section{Conclusion}
	
	In conclusion, we have experimentally demonstrated an advanced optical platform for quantitative dynamic phase mapping that eliminates the need for iterative phase retrieval algorithms. The advantages of this methodology stem from three core characteristics: (1) the \textbf{single-arm architecture} provides enhanced imaging robustness against environmental perturbations by eliminating the stringent stability requirements of traditional interferometric reference paths; (2) the \textbf{single-pixel detection scheme} effectively utilizes the broad temporal bandwidth of conventional digitizers, enabling high-speed continuous sampling to resolve rapidly evolving dynamics beyond the frame-rate bottlenecks of array sensors; and (3) the \textbf{computational field-correlation mechanism} enables direct, deterministic phase imaging, establishing a direct linear mapping from the heterodyne beat signal to the physical phase via intermediate frequency spectral analysis. By mathematically formulating the dynamically perturbed refractive index as a pure-phase object through a discrete matrix-based model, we achieve a direct extraction of the two-dimensional spatial phase distribution.
	
	While the system is validated here using ultrasonic waves, the acoustic field serves primarily as a high-frequency benchmark to demonstrate the system's spatiotemporal precision. Utilizing our phase-retrieval-free single-arm CD-GI architecture, we successfully reconstructed the dynamic spatial phase profiles of diverse complex geometries. The experimental results reveal that the optically extracted spatial wavelengths align closely with theoretical dispersion models, while the retrieved dynamic phase maintains a robust linear correlation with local physical modulation levels. By avoiding the convergence instabilities and phase ambiguities inherent to iterative retrieval algorithms, this direct linear methodology establishes a reliable, real-time-capable metrological tool. This robust framework offers a high-speed quantitative phase solution for characterizing rapidly evolving transparent phenomena, such as transient aeroacoustic flows, shockwave propagation, and the high-throughput dynamic analysis of biological cells in microfluidic environments.
	
	\begin{acknowledgments}
		The authors gratefully acknowledge funding support from the National Key R\&D Program of China (Grant No. 2024YFB4504002), the Strategic Priority Research Program of the Chinese Academy of Sciences (Grant No. XDB1690300),  the Space Application System of China Manned Space Program and the National Key R\&D Program of China (Grant No. JCKY2024110C034).
		
		C.W. and S.H. conceived the idea. C.W., J.Q., and S.L. performed the experimental measurements, X.J. performed the COMSOL numerical simulations, C.W. analyzed the experimental data. All authors discussed the results and contributed to the manuscript.
	\end{acknowledgments}
	
	\section*{Disclosures}
	The authors declare no competing interests.
	
	\section*{Data Availability}
	
	The data that support the findings of this study are available from the corresponding author upon reasonable request.
	
	%aipnum4-2.bst 2019-01-14 (MD) hand-edited version of apsrev4-1.bst

\end{document}